\documentclass[final,5p,times,twocolumn,lefttitle]{elsarticle}
\usepackage{amssymb}
\usepackage{lipsum}
\usepackage{amsmath}
\usepackage{xcolor}
\usepackage{color}
\usepackage{float} 
\usepackage{graphicx}
\usepackage{caption}
\usepackage{subcaption}
\usepackage{soul}
\usepackage[T1]{fontenc} 
\setcitestyle{square} 
\biboptions{comma}
\usepackage[colorlinks=true, linkcolor=red, citecolor=green, urlcolor=magenta]{hyperref}

\DeclareMathOperator{\arctanh}{arctanh}
\begin{document}

\begin{frontmatter}
\title{On entropic cosmology, late time acceleration and the quantum bounce.}

\author[First]{Luis Rey Diaz-Barron}
\ead{lrdiaz@ipn.mx}
\author[Second]{G. E. P\'erez-Cuéllar}
\ead{perezcg2015@licifug.ugto.mx}
\author[Second,Third]{M. Sabido}
\ead{msabido@fisica.ugto.mx}
\affiliation[First]{organization={Instituto Polit\'ecnico Nacional, Unidad Profesional Interdisciplinaria
	de Ingenier\'ia Campus Guanajuato.},
addressline={}, city={Silao},
            postcode={36275}, 
            state={Guanajuato},
            country={M\'exico}}
\affiliation[Second]{organization={Departamento de F\'isica de la Universidad de Guanajuato},
addressline={A.P. E-143}, city={Le\'on},    
            postcode={37150},
            state={Guanajuato},
            country={M\'exico}} 
\affiliation[Third]{organization={Department of Theoretical Physics, University of the Basque Country UPV/EHU},
addressline={P.O BOX 644}, city={},
            postcode={48080}, 
            state={Bilbao},
            country={Spain}}

\begin{abstract}

In this  work, we present a unified entropic description of the quantum bounce that is present in Loop Quantum Cosmology (LQG) and dark energy that drives the late time acceleration of the Universe. By considering a minimal area $A_0$ for the Universe and a modified entropy area constructed as the sum of $S_Q$, $S_{BH}$ and $S_{DE}$, we derive the Friedmann equations. We show that in the limit $A\sim A_0$, there is a critical density that generates the quantum bounce and  depends on the minimal area. Moreover, in the limit $A\gg A_0$ (large Universe) the late time dynamics gives an accelerating scale factor. Therefore, we can argue  an entropic origin for the quantum bounce and dark energy.
\end{abstract}
\end{frontmatter}

\section{Introduction} \label{Int}
{The formulation of General Relativity (GR) is one of the most outstanding achievements in theoretical physics. The current results from gravitational wave astronomy cement GR as the appropriate theory to describe the gravitational interaction. Although the open problems of dark energy and dark matter are compatible with GR (if one proposes new exotic sources of matter and energy), current observations do not discard alternative theories of gravity. Furthermore, considering that a complete quantum theory of gravity is still missing, one can  strongly consider the search for alternatives to GR. The usual approach is to consider gravity as a fundamental interaction, and from a fundamental principle write down the corresponding theory (i.e., $f(R)$ \cite{Sotiriou:2008rp,odintsov1}, massive gravity \cite{deRham:2014zqa}, Horndeski \cite{Horndeski:1974wa}, etc.).}
{An alternative to avoid  the inconsistencies of quantizing GR, is to consider gravity  as an emergent interaction. This approach is not new, the first foray in this direction was done by Jacobson \cite{Jac}, he derived Einstein's equations  from an entropy proportional to the area and in the process showed that  GR can be considered as an entropic force. More recently, there was a resurgence with Verlinde's ideas \cite{Ver}, where  Newtonian gravity is considered an entropic force 
that is derived from the entropy area relationship.} Other proposals make use of the holographic principle, for instance in \cite{vR}, invoking the laws of entanglement to derive Einstein equations. In these formulations, the fundamental expression is the Bekenstein-Hawking (BH) entropy.  
In this context  the modifications to the gravitational force are encoded in the entropy-area relationship. This has allowed to study the effects of noncommutativity, ungravity and non-extensive entropies in gravitational systems. One simply uses the modified BH entropy and either the modified Newtonian force or Friedmann equations, and study its phenomenological consequences \cite{Martinez-Merino:2017xzn,Perez-Cuellar:2024lau,Nicolini:2010nb}. To construct the modified entropy, one can propose under some general assumptions a functional form for the entropy. For example, if we want to consider a modification that can mimic dark energy, we know that the presence of a scalar field can drive an acceleration. Moreover, we know that in QFT, the degrees of freedom for the thermodynamical limit is achieved when  $N\to\infty$, $V\to\infty$ and $N/V$ finite, then one can argue that this behavior can be encoded on a term that scales with the volume and considering that the volume grows faster than the area, it should dominate in the late evolution of the Universe. Then by adding a term proportional to $A^{3/2}$ in the BH entropy, one finds an accelerating solution in the late time limit. In the the limit $t\to\infty$, the evolution equation allows to derive and effective cosmological constant, but it behaves as an evolving dark energy. Moreover, this type of volumetric term, can be used to model galactic rotation curves.\\
Then one should start by introducing some assumptions to propose a particular functional form of the entropy. As we are expecting to obtain a modified theory of gravity, the entropy will be a function of the area $A$, and in some limit a term proportional to area should appear. Also, as we want to eliminate singularities at the origin then is seems plausible that there is a minimum area $A_0$. The concept of a minimum scale, can be justified from inherent minimum scale in a quantum theory of gravity, i.e in loop quantum gravity a discrete spectrum for the area operator and minimum value for the area is predicted \cite{DePietri:1996tvo}. As this result is of quantum origin, one would expect that for an area much larger than the minimum area,  one should recover an entropy in which the usual Bekenstein-Hawking term dominates, therefore by modifying the entropy as
\begin{equation}
S=\frac{A}{\kappa}\sqrt{1-\frac{A_0}{A}},
\end{equation}
where $\kappa$ is a constant to be be fixed.  After taking a series expansion in $A_0\ll A$ one recovers to first order the usual BH entropy $S_{HB}$ by setting $\kappa=4G$. In the study of quantum black holes, there is strong evidence that the quantum effects induces a correction to the entropy that is proportional to the logarithm of the area \cite{Gour:1999ta,Obregon:2000zd,Meissner:2004ju,Ghosh:2004rq}. This is a contribution that arises in a semiclassical approximation, consequently one can regard the logarithmic correction as being of quantum origin. Therefore, if we want to account for quantum aspect of gravity,  one expects that the entropy includes a term 
{ that is related} to the logarithm of the area. Considering the previous points, one can propose the following entropy
\begin{equation}\label{qbounce}
S(A)= \alpha S_Q\left(\sqrt{1-\frac{A_0}{A}}\right) + \frac{A}{4G}\sqrt{1-\frac{A_0}{A}},
\end{equation}
where $\alpha$ is a constant and  $S_Q$ is a function that we expect to describe the quantum regime and should be fixed by connecting to a particular model of quantum gravity. The remaining question is what physical model should we explore? Although we could try to study black holes, we will focus our attention in quantum cosmology or, more precisely, Loop Quantum Cosmology (LQC) \cite{Bojowald:2005epg,Ashtekar:2006wn,Ashtekar:2008gn}. As we will be interested in recovering\footnote{Reproducing the quantum bounce from the modified Friedmann equations fixes the value $\alpha=\frac{A_o}{4G}$.} the equations of LQC, we will follow the approach in \cite{Cai:2005ra}, where studying the thermodynamics at the apparent horizon leads to the Friedmann equations from the entropy-area relationship.\\
As we will be considering the cosmological scenario, we can broaden our search to see if we can account for dark energy. As already commented, it is possible to induce a late time acceleration on the scale factor and define an effective cosmological constant. Therefore, we can model the dark energy sector by adding an extra term to the entropy area relationship. Therefore, we can propose the following
\begin{align}\label{qbounce_dark}
&S(A)=\\
&\frac{A_0}{4G} S_Q\left(\sqrt{1-\frac{A_0}{A}}\right) + \frac{A}{4G}\sqrt{1-\frac{A_0}{A}}+\epsilon~ S_{DE}\left(\frac{A}{4G}\sqrt{1-\frac{A_0}{A}}\right).\nonumber
\end{align}
The $\sqrt{1-\frac{A_0}{A}}$ discriminates between the scales $A\sim A_0$ where the quantum effects should dominate and  $A\gg A_0$ (late time or large Universe) where one expects an accelerating scale factor driven by an effective dark energy
.\\
The main goal of this work is to construct an entropic cosmological model where one can incorporate both the quantum bounce  from the hypothesis of a minimal area $A_0$ and for large $A$ (or equivalently late time) gives an accelerating scale factor without including dark energy or a cosmological constant. Using some simple physical requirements, we construct a modified entropy-area relationship and analyzing the thermodynamics at the apparent horizon derive the Friedmann equations that in the appropriate limit, reproduce the quantum bounce and the effective cosmological constant. \\
This work is organized as follows. In Sec.~\ref{bounce} we propose a particular form for the modified entropy-area relationship and following the procedure in \cite{Cai:2005ra}, show that one can have the quantum bounce in therms of the minimal area $A_0$. In Sec.~\ref{dark} we include a term that for  $A\gg A_0$ has a volumetric contribution to the entropy and verify for the Friedmann equations, that it not only gives and effective cosmological constant, but also reproduces the quantum bounce. Lastly, Sec.~\ref{conclusions} is devoted for final remarks.
\section{Entropic Origin for the Quantum Bounce}\label{bounce}

Before turning our attention to the quantum bounce, we will briefly {show} the derivation of the Friedmann equations from the Hawking-Bekenstein entropy $S_{HB}$ and thermodynamics at the apparent horizon\cite{Cai:2005ra}, as this will be the method to derive the dynamics of the Universe from the modified entropy-area relationship. Starting with a Friedmann-Robertson-Walker metric
\begin{equation}
    d s^2=-d t^2+a^2\left(\frac{d r^2}{1-k r^2}+r^2 d \Omega^2\right)=h_{a b} d x^a d x^b+\tilde{r}^2 d \Omega^2,
\end{equation}
where $a$ is the scale factor and the dynamical radius is given by $\tilde{r} = a(t)r$. The apparent horizon, which is also a dynamical quantity, satisfies the equation $h^{a b} \partial_a \tilde{r} \partial_b \tilde{r}=0$ and  is given by
\begin{equation}
    \tilde{r}_A=\frac{1}{\sqrt{H^2+k / a^2}},
\end{equation}
in terms of the Hubble parameter $H = \frac{\dot{a}}{a}$. Another relation is obtained by differentiating 
\begin{equation}
    \dot{\tilde{r}}_A=-H \tilde{r}_A^3\left(\dot{H}-k / a^2\right).
    \label{eq:dothorizon}
\end{equation}
From the energy momentum tensor of a perfect fluid, the energy conservation gives the continuity equation $\dot{\rho}=-3 H(\rho+p)$.
Other useful quantities are the work density $\omega=(1 / 2) T^{a b} h_{a b}$ and energy supply vector $\psi_a=T_a^b \partial_b \tilde{r}+\omega \partial_a \tilde{r}$ 
which for the FRW case take the form
\begin{equation}
    \omega=1 / 2(\rho-p),~\psi_a=-1 / 2(\rho+p) H \tilde{r} d t+1 / 2(\rho+p) a d r.
\end{equation}
Now a relation for the energy flux passing trough the apparent horizon in an infinitesimal time interval can be constructed as
\begin{equation}
    \delta Q=-A \psi=A(\rho+p) H \tilde{r}_A d t
\end{equation}
where $A$ represents the apparent horizon area. We also assume 
that the apparent horizon has a temperature and entropy given by the Bekenstein-Hawking entropy $S_{HB}=A/4G$ and Hawking temperature\footnote{We are using units where $c = k_B = \hbar = 1$.} $T=1/{(2 \pi \widetilde{r}_A})$.
Writing the entropy in terms of the apparent horizon radius $\tilde{r}_A$ and using Eq.(\ref{eq:dothorizon}), the differential of the entropy can be written as
\begin{equation}
    d S=-\frac{2 \pi}{G} \tilde{r}_A^4 H\left(\dot{H}-\frac{k}{a^2}\right) d t.
\end{equation}
From the Clausius relation $\delta Q=T d S$,
it follows that $A(\rho+\rho) H \tilde{r}_A d t = TdS$, therefore
\begin{equation}
    \dot{H}-\frac{k}{a^2}=-4 \pi G(\rho+p),
\end{equation}
using continuity equation 
\begin{equation}
    H\left(\dot{H}-\frac{k}{a^2}\right)=\frac{4 \pi G}{3} \dot{\rho},
\end{equation}
and after integrating the previous expression, we recover the Friedmann equation\footnote{We have used $\frac{1}{2} \frac{d}{d t}\left(H^2+\frac{k}{a^2}\right)=H\left(\dot{H}-\frac{k}{a^2}\right)
$.}
\begin{equation}
    H^2+\frac{k}{a^2}=\frac{8 \pi G}{3} \rho.
\end{equation}
Now we  explore the quantum regime and as first step we will only use the first two terms in Eq.(\ref{qbounce_dark}) and propose a function related to the logarithm,  $S_Q(x)=\arctanh\left(x\right)$.  Then the modified entropy-area relationship is
\begin{equation}
{S=\frac{A_0}{4G}\arctanh\left(\sqrt{1-\frac{A_0}{A}}\right)+\frac{A}{4G}\sqrt{1-\frac{A_0}{A}}},
\end{equation}
after taking the differential of entropy 
and the Clausius relation $\delta Q = TdS$ we have
\begin{equation}
    A(\rho+p) H \tilde{r}_A=\frac{8 \pi \tilde{r}_A \dot{\tilde{r}}_A}{8 \pi G \tilde{r}_A \sqrt{1-\frac{A_0}{4 \pi \tilde{r}_A^2}}}.
\end{equation}
Using the relations for the horizon radius 
and  the continuity equation we get
\begin{equation}
    \frac{4 \pi G}{3} \dot{\rho}=\frac{1}{\sqrt{1-\frac{A_0}{4 \pi}\left(H^2+\frac{k}{a^2}\right)}} \frac{1}{2} \frac{d}{d t}\left(H^2+\frac{k}{a^2}\right),
\end{equation}
and after integration we arrive to 
\begin{equation}
    \frac{4 \pi G}{3} \rho= -\frac{4 \pi}{A_0} \sqrt{1-\frac{A_0}{4 \pi}\left( H^2+\frac{k}{a^2}\right)}+\frac{4 \pi}{A_0}.
\end{equation}
Finally, solving for $H^2+\frac{k}{a^2}$, we get
\begin{equation}\label{LQC}
    H^2+\frac{k}{a^2}=\frac{8 \pi G}{3} \rho\left(1-\frac{A_0 G}{6}\rho\right)=\frac{8 \pi G}{3} \rho\left(1-\frac{\rho}{\rho_c}\right),
\end{equation}
after identifying the critical density $\rho_c \equiv 6/A_0 G$, one recovers the LQG Friedmann equation that describes the quantum bounce, with the critical density in terms of the minimal area.

\section{Entropic Origin for Dark Energy}\label{dark}
We now turn our attention to the full model where the proposed entropy-area relationship has to encode both the quantum bounce and the late time acceleration of the Universe. As previously discussed, the inclusion of a volumetric term $S\sim A^{3/2}$ induces an accelerating scale factor for the late time  dynamics of the Universe\cite{Diaz-Saldana:2018gxx,Chagoya:2023hjw}. Moreover, the $S_{DE}$ term should be overshadow by the first two terms in Eq.(\ref{qbounce_dark}) when $A\sim A_0$. Also, when $A\gg A_0$ the term $S_{DE}$ should dominate and the resulting Friedmann equations must describe and accelerating Universe. The simplest entropy-area relationship that satisfies these requirements is
\begin{align}\label{entropy_dark}
S&=\frac{A_0}{4G}\arctanh\left(\sqrt{1-\frac{A_0}{A}}\right)
+\frac{A}{4G}\sqrt{1-\frac{A_0}{A}}\\
&+\epsilon\left({\frac{A}{4G}}\sqrt{1-\frac{A_0}{A}}~\right)^{3/2}.\nonumber
\end{align}  
As before we start from the Clausius relation $\delta Q = TdS$, we consider the continuity equation, and  using $A = \frac{4\pi}{\left(H^2 + \frac{k}{a^2}\right)}$, we get
\begin{align}\label{eq_master}
    \frac{8\pi G}{3}\rho &= 4G \int S^\prime\left(A\right) d\left(\frac{4\pi}{A}\right)\\
    &=4G \int S^\prime\left(\frac{4\pi}{H^2 + \frac{k}{a^2}}\right) d\left(H^2 + \frac{k}{a^2}\right),\nonumber
\end{align}
{where $S^\prime(A) = \frac{dS}{dA}$}. To obtain the Friedmann equations, starting from this equation\footnote{The cases discussed are also obtained from this equation.} we simply solve for $H^2 + \frac{k}{a^2}$.\\
Using the entropy Eq.(\ref{entropy_dark}) 
\begin{align}
    &\frac{8 \pi G}{3} \rho=\\
    &4 G \int\left[\frac{1}{4 G} \frac{1}{\sqrt{1-\frac{A_0}{A}}}+\frac{3 \epsilon}{16 G^{3 / 2}} \frac{A^{1 / 2}}{(1-\frac{A_0}{A})^{1 / 4}}\left(1-\frac{A_0}{{2}A}\right)\right] d\left(\frac{4 \pi}{A}\right),\nonumber
\end{align}
after the change of variable  $x^2 = 1 - \frac{A_0}{A}$  
Eq.(\ref{eq_master}) takes the form 
\begin{align}
    \frac{8\pi G}{3}\rho &=-\frac{8 \pi}{A_0}\int dx {- \frac{3 \epsilon{\pi}}{\sqrt{G A_0}} \int \frac{x^{1 / 2}}{\sqrt{1-x^2}} d x}\\
    &- \frac{{3} \epsilon {\pi}}{\sqrt{G A_0}} \int \frac{x^{5 / 2}}{\sqrt{1-x^2}} d x,\nonumber
\end{align}
and finally we arrive at
\begin{align}
    \frac{8\pi G}{3}\rho &=-\frac{8 \pi}{A_0} x-\frac{{2}\epsilon {\pi}}{5\sqrt{G A_0}} x^{3 / 2}\\
    &\times\left[-{3}\sqrt{1-x^2}+{8}{ }_2 F_1\left(\frac{1}{2}, \frac{3}{4}, \frac{7}{4} ; x^2\right)\right]+C.\nonumber
\end{align}
Unlike the previous cases where one can solve for $\left( H^2+\kappa/a^2\right)$, we get a transcendental equation, but even if we can not solve for the Hubble parameter we can explore different limits to obtain the corresponding Friedmann equation. Moreover, as we are interested in the quantum regime of the Universe as well as the late time dynamics, in terms of the area, these two regimes are given by $A\sim A_0$ and $\frac{A_0}{A}\ll1$, respectively, and therefore we will focus our attention to these limits.

First we verify the presence of the bounce, for this we do an expansion to first order in $x$  on the modified Friedmann equation 
\begin{equation}
    \frac{8 \pi G}{3} \rho=-\frac{8 \pi}{A_0} \sqrt{1-\frac{A_0}{4 \pi} \left[H^2+\frac{k}{a^2}\right]}+C,
\end{equation}
solving for $H^2 + \frac{\kappa}{a^2}$ we get
\begin{equation}
    H^2+\frac{\kappa}{a^2}=\frac{8\pi G}{3}\rho\left(\frac{A_0 C}{8\pi}-\frac{A_0G}{6}\rho\right)+\frac{4\pi}{A_0}-\frac{A_0}{16\pi}C^2,
\end{equation}
and after fixing the integrating constant $C=\frac{8\pi}{A_0}$ {and the critical density} $\rho_c=\frac{6}{A_0G}$, the modified Friedmann equation reduces to Eq.(\ref{LQC})
recovering the Friedman equation from LQC. \\
The next step is to analyze the  late time behavior, this corresponds to a large Universe, $A\gg A_0$, and therefore is consistent with small\footnote{For  $A\gg A_0$,  means  that the critical density $\rho_c \gg \rho$, implying that $A \to \infty$.} $\rho$. The Friedmann equation derived from Eq.(\ref{entropy_dark}) and Eq.(\ref{eq_master}), in the limit of large area is
\begin{equation}\label{despejado}
    H^2+\frac{k}{a^2}=\frac{8 \pi G}{3} \rho+\frac{9}{2} \frac{\pi}{G} \epsilon^2 {+} \frac{1}{2} \sqrt{\frac{81 \epsilon^4 \pi^2}{G^2}+96 \pi^2 \epsilon^2 \rho},
\end{equation}
which in  the limit $\rho\to0$, 
gives an effective cosmological constant\footnote{We have used the same value for C.}
\begin{equation}
    \Lambda_{eff}=\frac{9\pi}{G}\epsilon^2.
\end{equation}
{This late-time  behavior is driven by the volumetric correction term in Eq.(\ref{entropy_dark}). This is the same result one gets by including the volumetric term $A^{3/2}$ into the usual Bekenstein-Hawking entropy \cite{Diaz-Saldana:2018gxx,Chagoya:2023hjw}.\\
To better understand the dynamics for large $A$, we follow \cite{Chagoya:2023hjw}, first we define {the densities} for the terms  associated to the parameters $\epsilon$, matter and spatial curvature 
\begin{equation}
\Omega_{\epsilon 0}\equiv\frac{1}{2H_{0}^2}\frac{{9\pi}}{G}\epsilon^{2},\quad\Omega_{M 0}=\frac{8\pi G}{3H_{0}^{2}}\rho_{0},\quad \Omega_{\kappa 0}=-\frac{\kappa}{a_{0}^2H_{0}^2}~.
\end{equation}
After considering non-relativistic matter for which $\rho=\rho_0 (1+z)^3$, where $z$ is the redshift  $1+z=1/a$, Eq.~(\ref{despejado}) can be written in terms of the above parameters as
\begin{equation}\label{Hentropic}
H^{2}(z)=H_{0}^{2}\left\{ \Omega_{\kappa 0}(1+z)^{2}+ \frac{1}{2}\left[\sqrt{\Omega_{\epsilon 0}}+\sqrt{\Omega_{\epsilon 0}+2\Omega_{M 0}(1+z)^{3}}\right]^{2} \right\},
\end{equation}
and the modified Friedmann constraint is
\begin{equation}\label{omegaentropic}
\Omega_{\kappa 0}+\frac{1}{2}\left(\sqrt{\Omega_{\epsilon 0}}+\sqrt{\Omega_{\epsilon 0}+2\Omega_{M 0}}\right)^{2}=1.
\end{equation}
Interestingly it is equivalent to 
\begin{equation}
H^2(z)=H_0^2\left\{ \Omega_{\kappa 0}(1+z)^2+\left[ \sqrt{\Omega_{r_c 0}}+\sqrt{\Omega_{r_c 0}  +\Omega_{M 0}(1+z)^3}\right]^2 \right \}.
\end{equation}
where $r_c=\frac{M^{2}_{pl}}{2M^3_{(5)}}$ and the densities are given by
\begin{equation}
\Omega_{\alpha 0}=\frac{\rho^0_\alpha}{3M^2_{pl}H^2_0a_0^{3(1+w_\alpha)}},~ \Omega_{\kappa 0}=-\frac{\kappa}{H^2_0a_0^{2}},~ \Omega_{r_{c} 0}=\frac{1}{4r^2_cH^2_0}.
\end{equation}
Consequently,  for $A\to \infty$ this model has the same dynamics as the DGP model.

\section{Final Remarks}\label{conclusions}
In this work we present a model constructed from a modified entropy-area relationship,
that allowed us to give an entropic origin to both  the quantum bounce of LQC and to dark energy.
This is achieved by proposing a minimal area $A_0$, (which implies a minimum size of the Universe) in conjunction with  a modified entropy-area relationship. The modified entropy is the the sum of three terms, one related to the quantum regime, one related to the Hawking-Bekenstein entropy and a term related to dark energy. Additionally, the decoupling of the quantum regime (early-time)  and late-time cosmological dynamics of the universe, is determined by the factor $\sqrt{1-A_0/A}$, for this reason all terms in the entropy are to be multiplied by this factor. 
Moreover, for large $A$ the entropy is dominated by the volumetric term and we expect to have an accelerating Universe, and for $A\sim A_0$  the first two terms dominate and would expect the bounce. But even if one has the correct  behavior on the entropy, it  does not necessarily mean that this will be reflected in the Friedmann equations. Particularly, that the  quantum bounce will be recovered once we include  the volumetric term on the entropy. \\
To study the cosmological implications of these ideas, we derive the Friedmann equations from the thermodynamics at the apparent horizon. The resulting  equations are complicated and give a transcendental function for the Hubble parameter, for this reason we had to  analyze the dynamics in different  limits, mainly $A\sim A_0$ and $A\gg A_0$. For $A\sim A_0$, we obtain the Friedmann equations that reproduce the quantum bounce of LQC and we find that the critical density is a function of the minimal area $A_0$. Therefore, we can conclude that although we maintained the volumetric term on the entropy, those degrees of freedom decouple in such a manner that the bounce is maintained. For large $A$, we recover the Friedmann equations that in the limit $A\gg A_0$ is consistent with an accelerating scale factor and in the limit $t\to \infty$ gives an  effective cosmological constant. As expected, in this limit where the entropy is dominated by the volumetric term, we find that the model is equivalent to the DGP model. Consequently, we conclude the following: by postulating a minimum area $A_0$ for the Universe and a volumetric contribution to the entropy area relation (the factor $\sqrt{1-A_0/A}$ multiplying the area $A$) one recovers both the quantum bounce and the accelerating scale factor.
{To explore the the phenomenology of the model in the regime between the two limits discussed, we can perform a series expansion and derive the Friedmann equation as a power series in the Hubble parameter. This approach was adopted in \cite{Chagoya:2024tqv}, where the authors express the entropy as power series in the area and derived the corresponding Friedman equations as a series in the Hubble parameter to compare with observations\footnote{The case $A>>A_0$ corresponds to the one in \cite{Chagoya:2024tqv}, if we take $N=2$, $\alpha=\frac{A_0}{4G},\sigma_0=\frac{A_0\epsilon}{16\pi},\sigma_1=0,\sigma_2=\epsilon$.}.  
This line of research is currently underway and will {be} reported elsewhere.}

In summary, the use of thermodynamics at the apparent Horizon allows to systematically
introduce physical hypothesis to explain different aspects of the Universe. In particular, we have presented a model that accounts for the quantum bounce and the current acceleration of the Universe by introducing a minimal area and corrections to the entropy area relationship, allowing us to simultaneously  give an entropic origin for  both the quantum bounce and dark energy.

\section*{Acknowledgements}
{\bf M.S.} is supported by the grant CIIC 027/2026 and by the program ``Estancias Sab\'aticas para la consolidaci\'on de grupos de investigaci\'on'', SECIHTI and by SNII-SECIHTI. {\bf G.E.P-C.} is supported by the program ``Becas Nacionales de Posgrado'', SECIHTI. {\bf L.R.D.B.} was partially supported by SNII-SECIHTI and Secretaría de Investigación y Posgrado del Instituto Politécnico Nacional grant IND-2026-0041. 
\bibliographystyle{unsrt}
\bibliography{bib}
\end{document}